\documentclass[12pt, preprint]{aastex}

\begin{document}
\title{An axisymmetric, hydrodynamical model for the torus wind in AGN.}

\author{A. Dorodnitsyn\altaffilmark{1,2}, T. Kallman\altaffilmark{1}, and 
D. Proga\altaffilmark{3}}

\altaffiltext{1}{Laboratory for High Energy Astrophysics, NASA Goddard Space Flight Center, Code 662, Greenbelt, MD, 20771, USA}
\altaffiltext{2}{Space Research Institute, Profsoyuznaya st., 84/32, 117997, Moscow, Russia}
\altaffiltext{3}{Department of Physics and Astronomy, University of Nevada, Las Vegas, NV 89154, USA}
\begin{abstract}
We report on time-dependent axisymmetric simulations of an X-ray excited
flow from a parsec-scale, rotating, cold torus around an active galactic
nucleus. Our simulations account for radiative heating and cooling and
radiation pressure force. The simulations follow the development of
a broad bi-conical outflow induced mainly by X-ray heating.
We compute synthetic spectra predicted by our simulations.
The wind characteristics and the spectra
support the hypothesis that a rotationally supported torus
can serve as the source of a wind which is responsible for the warm
absorber gas observed in the X-ray spectra of many Seyfert galaxies.
\end{abstract}

\section{Introduction}

Approximately $50\%$ of low-red-shift AGNs exhibit spectra rich 
in atomic absorption features in the X-ray band. These spectra, referred to as warm absorbers, contain numerous lines and bound-free edges from ions of 
intermediate-Z elements, and indicate blue-shifts of the absorbing material of $\simeq 10^2 - 10^3$ km s$^{-1}$.   
The term refers to the fact that the X-ray absorbing gas has an electron temperature and ionization degree which 
are intermediate between that expected for neutral and fully ionized material.
Warm absorber gas has been known for more than two decades,  beginning with observations using the  {\it Einstein} Observatory satellite \cite{Halpern84}, although 
the limited spectral resolution of early X-ray experiments hampered a detailed study of its properties.

Recent X-ray telescopes, {\it Chandra} and XMM-{\it Newton}, provide unprecedented spectral resolution up to $\sim$10 keV,
and have revealed much of the complexity of warm absorber spectra.
The ubiquity of X-ray warm absorbers and the implication of outflow in objects which are likely powered by accretion, 
combine to make warm absorbers of great interest for the 
understanding of active galactic nuclei.

While observationally the reality of the X-ray absorbing gas is firmly established, the source of this gas remains uncertain.
If the wind originates from an accretion disk 
\newline \cite{Konigl94,Murray95},  
then warm absorber gas can occupy a vast spatial region from the inner accretion disk up to and 
beyond the broad line region.  In addition, it is likely to have a speed which is comparable to 
the escape velocity in the region where the wind originates. If so, the line widths favor a wind originating 
far outside the inner accretion disk, at  $\sim 1\, {\rm pc}$, for an AGN containing a $10^6 M_\odot$ black hole. 

The natural source of gas at this distance is the molecular torus which is responsible for obscuring the 
broad line region in Seyfert 2 galaxies, and which is thought to exist in some form in many low and 
intermediate luminosity AGN \cite{AntonucciMiller86}.
The existence of an outflow from the torus has been suggested by \cite{KrolikBegelman86, KrolikBegelman88}, 
and as the source of warm absorber flows by 
\cite{KrolikKriss1,KrolikKriss2}.

Direct interferometric observations of this dusty structure have recently
become available: mid-infrared high spatial resolution studies of the active
nucleus of NGC 1068,  revealed a dust torus - like structure that is 2.1 pc thick
and 3.4 pc in diameter ~\cite{JaffeNATUR}. Evidence of at least two temperature components has been established: the first one is extended and warm (320 K) and
conceals the second, which is compact and hot ($>800 K$). It is natural to assume that the latter component can be 
attributed to the inner part of the torus, heated by the radiation from the
compact nucleus.  
Although such evidence is most solid for nearby
active galaxies it is likely that the same
obscuring torus paradigm may apply in quasars whose central regions are
heavily obscured by gas and dust (Type II quasars)
\cite{Zakamska06}. 

The work done so far in modeling torus warm absorber flows employ either one-dimensional models e.g. \cite{ChelNet05} or two dimensional models which omit rotational forces and also employ 
an artificial shape for the torus, which is adopted as a boundary for the flow \cite{BalsKrolik}.
These calculations support the idea that the outflow originates  at $r\sim 1\,$ pc - the estimate supported both from 
virial arguments and also deduced from the warm absorber line widths.  
The content of the warm absorber gas is highly ionized cosmically abundant  heavy elements. This gas can occupy a region from 0.01 pc to $\sim\,$10 pc from the nucleus. The column densities are in the range $10^{21} - 10^{24} \, {\rm cm}^{-2}$.
These estimates are consistent with constraints on the position of the warm absorber gas, which are deduced from
the absence of correlated variability of the line equivalent widths to changes of the continuum \cite{Net03, Beh03}.  

The goal of the current work is to show that  a state-of-the-art  hydrodynamic model of the evaporative  torus wind 
possesses key characteristics of the the warm absorber.  
To approach this goal, in this paper we present results of models containing the following ingredients:
1) Time-dependent axisymmentric numerical simulations that take into account
rotational forces , X-ray heating and cooling 
and radiation pressure both in spectral lines and continuum;
2) Exploration of the resulting distributions of density, velocity etc.; and
3) A calculation of the synthetic X-ray absorption spectra using the density 
and velocity distributions as input. Although calculating the spectra is crucial to check the relevance of the
particular hydrodynamic solution to the warm absorber phenomenon,  the primary goal of the work reported 
here is the outflow dynamics .

\section{Basic equations}

\noindent
The equations to be solved include:

\begin{eqnarray}
\frac{\partial\rho}{\partial t}+\nabla \cdot (\rho {\bf v})=0\mbox{,}\label{eqCont}\\
\rho \left(\frac{\partial {\bf v}} {\partial t}+({\bf v}\cdot\nabla){\bf v}\right)
=-\nabla p-\rho\nabla\Phi+\rho\,{\bf g_{rad} } \mbox{,}\label{eqMoment}\\
\frac{\partial\epsilon}{\partial t}+{\bf \nabla}\cdot\left({\bf v}(\epsilon+p+\rho\Phi)\right) =
H\mbox{,}\label{eqEnergy}
\end{eqnarray}
these are the conservation equations for: mass, momentum and energy. Heating and cooling processes are 
described by the function $H({\rm erg\, cm^{-3}\, s^{-1} }) $; $\epsilon$ - is the sum of the kinetic and internal energy densities:
$\epsilon=\rho \, v^2/2+e$. Three components of the velocity are taken 
into account. However, the adopted azimuthal symmetry implies that all 
$\partial/\partial\phi=0$.
The gas is assumed to have a constant adiabatic
index, so that $e = P / (\gamma - 1)$ with $\gamma = 5/3$.
Equations (\ref{eqCont})-(\ref{eqEnergy}) are cast in a non - dimensional form with the characteristic scales set by the properties of the plasma orbiting at 
a distance of $R_0=1\,{\rm pc}$ from a black hole with mass $M_6$ in units of $10^6 M_\odot$: 
time in measured units of $t_0({\rm s})=4.7\cdot10^{11}\,r_{\rm pc}^{3/2}\,M_6^{-1/2}$, velocity in units of 
$V_0({\rm cm\,s^{-1} })=6.6\cdot 10^6 M_6^{1/2} \, r_{\rm pc}^{-1/2}$. 

We anticipate that the conditions at the interface of a torus and an outflow driven by X-ray heating will resemble the X-ray 
self-excited wind in X-ray binaries.
That is, an X-ray heated skin is created at the outer layers of the companion star atmosphere as a result of
X-ray heating by the compact object. The temperature there rises almost discontinuously to some value that allows 
the gas to escape the gravity of the star. The flow below the jump is subsonic and is determined by the details 
of the heating, cooling, and transfer processes in this region \cite{Basko77}.  
In such a situation, 
the distribution of temperature 
depends on the heating and cooling, including both adiabatic effects and radiation, in the equations of 
hydrodynamics.  This can only be calculated accurately if the interior of the star is included as part of the 
computational domain, and the same applies to the problem of interest to us: the torus X-ray excited wind.
We, therefore include in our computational domain the entire volume containing the torus and 
we start our simulations from a rotating torus configuration in equilibrium 
in the external gravitational field of the black hole.  

A stationary, polytropic,  toroidal structure with constant angular momentum was suggested by 
\cite{PP84}(PP-torus), and was shown to be stable in 2.5 dimensions against 
radial or axial perturbations,  an found to be unstable to {\it non-axisymmetric} perturbations (this effect cannot
be numerically investigated in axisymmetry). 
In our dimensionless units the locus of the maximum torus density $\rho_{\rm max}$ occurs at $R_0$, so that $\rho_{\rm max}=1$.
The distortion of the torus is controlled by the parameter
$d=(\varpi^++\varpi^-)/(2R_0)$, where  $\varpi^\pm$ refers to the inner or outer edge of the torus.
No replenishing of the torus gas is assumed, i.e. gas can only leave computational domain.
The drawback of this approach is that it is not possible to obtain a true stationary solution of 
the problem.
Nonetheless, as we will show, the torus flow can be regarded as a quasi-equlibrium on timescales
shorter than the depletion timescale, and can thus provide representative snapshots of the 
character of the flow.

\subsection{X-ray exited wind:  evolution of the wind and the torus}

The thermodynamic properties of X-ray heated gas depend on the spectrum of the incident radiation as
well as on the local atomic physics. 
Under the assumption of photo-ionization
equilibrium the thermodynamic state of the gas
can be parameterized in terms of ratio of radiation energy density to baryon density \cite{Tarter69}:  
$\xi=4\,\pi\,F_{\rm x}/n\mbox{,}$
where $F_{\rm x}=L_{\rm x}e^{-\tau}/(4\pi r^2)$ is the local X-ray  flux, $L_{\rm x}$ is the X-ray luminosity of the nuclei, ${\displaystyle \tau=\int_0^r\kappa_e\rho\,dr}$ -is the optical depth. We assume that the attenuation is dominated by Thomson scattering
 $\kappa=0.2(1+X_{\rm H})\simeq0.4\,{\rm cm^{2}\,g^{-1}}$, where $X_{\rm H}$ is the mass fraction of hydrogen,  and the factor $e^{-\tau}$
- describes approximately the attenuation of the radiation on the way from the source toward a fiducial point. 
Assuming that the there is a fraction $\eta_x$ of the total luminosity $L_{\rm BH}$ available in X-rays  and that the disk radiates a fraction
$\eta_{edd}$ of its Eddington luminosity $L_{edd}=1.25\cdot 10^{44}\,M_6$ 
we estimate:  $\xi\simeq 4\cdot 10^2 \cdot\eta_x\,\eta_{edd}\, M_6/ (N_{23}\,r_{pc})$, where $r_{pc}$ -is the distance in parcecs and 
$N_{23}$ is the column density in $10^{23}$ ${\rm cm}^{-2}$.
If the wind is driven by thermal evaporative mechanism, then its pressure and temperature
are determined by the condition of photo-ionization equilibrium, if the dynamical time within the flow is much larger than the  characteristic time of the heating-cooling.
In a stationary flow the rate of the Compton and photo-ionization heating should be balanced by the Compton, radiative recombination, bremsstrahlung and line cooling respectively. In the calculations we adopt approximate formulas of \cite{Blondin94} for these processes. The radiative acceleration, ${\bf g_{\rm rad}}$, accounts for both the acceleration due to lines $g_{\rm rad}=(F_{\rm UV}\kappa/c)\,k\,t^{-\alpha} 
\left( 
(1+\tau_{\rm max} )^{(1-\alpha)}
-1 \right)
/\tau_{\rm max}^{1-\alpha}$, where $L_{\rm UV}$ is the UV flux, $t=\sigma_e\,v_{th}\rho/|dv/dr|$, $v_{th}$ is the thermal velocity, $\sigma_e$ is the electron-scattering opacity, and $\alpha=0.5$
\cite{CAK75, StevensKallman}, and for radiation pressure
on electrons $g_{\rm cont}=F_{\rm UV}\kappa/c$. 
In our approach, the radiation force which is due to lines and also $k$ and $\tau_{\rm max}$ depend on the ionization parameter $\xi$, and we adopt
the description made by \cite{StevensKallman}. 
The ionization parameter $\xi$ is calculated self-consistently at each grid zone, at each time step taking into account the attenuation of the X-ray flux due to finite {\it radial} optical depth.

We adopt spherical-polar $(r,\theta, \phi)$ coordinates for calculation of a wind  heated by a central source of radiation. 
We extend the computational domain ${r_i}$ from $r_{\rm in}=0.01$ pc
to $r_{\rm out}=50$ pc and the polar domain {${\theta_i}$} from $0$ to $\pi$ with no assumption about the equatorial symmetry. 
Models presented here have been calculated with a  $N_r\times N_\theta=100\times100$ grid. We have also performed test calculations on the $300\times300$
grid finding no noticeable difference with the result presented here. 
We postpone the discussion of this and related issues to a separate publication.  
Both angle and radial  grids are 
non-uniformly spaced: the angle gradually increases towards $\theta=\pi/4$ with fewer angular points towards the pole ($\theta=0$) and equator 
($\theta=\pi/2$), and analogously spaced in the southern hemisphere. The use of a logarithmically spaced radial grid allows for the "throat" of the torus to be resolved.  
We set outflowing boundary conditions at both ends of  the radial grid and axially symmetric at $\theta=0, \, \pi$. 
This initial torus is imbedded in a low density gas with $\rho=10^{-4}\rho_{max}$, where $\rho_{max}$ is the maximum density inside the initial torus.
To solve the hydrodynamical part of the system of equations (\ref{eqCont})-(\ref{eqEnergy}) we use the ZEUS-2D code
described by \cite{StoneNorman92}. 
As a test we have evolved a torus distribution of matter for $t=2$ (two rotational periods in our units) and found the whole
configuration stable. 
We include the heating term into the hydrodynamical equations using a fully implicit approach because the characteristic heating time scale can be much shorter than the dynamical time. 

We assume $M_6=1$,  $\eta_{edd}=0.4$ , $\eta_{\rm x}=0.5$ and also that $0.5\,L_{\rm BH}$ is radiated in UV and contributes to the flow energetics through the radiation force ${\bf g_{\rm rad}}$. 
The initial torus configuration chosen here is only mildly distorted, the distortion parameter: $d=1.25$. 
Following \cite{PP84}, we assume that the distribution of specific
angular  momentum inside the initial torus, is constant.
To ensure that our initial torus is in equilibrium, we follow \cite{PP84} with the exception that we reduce the effective gravity $GM_{\rm BH}/r^2(1-\eta_{\rm edd})$ due to the effect of the radiation pressure. However the real continuum radiation pressure, which is adopted in the calculations, is a factor of 
$e^{-\tau_r}$ less because of the optical depth effect.
Our initial torus is Thomson thick $\tau_r(\varpi^+,\theta=\pi/2)\simeq 4.35$.
Although the torus itself is not the primary topic of this study, the X-ray illumination does affect the interior and average properties of the torus.
During the evolution the locus of the maximum torus density (located initially at 
$\varpi=1$) shifts towards larger $\varpi$. This happens primarily because of the over-pressured region created by the evaporating gas and also because of radiation pressure.
 This allows for the energy to be deposited directly to the torus interior. 
In our case, direct radiative heating produces gradients of {\it gas} pressure in the torus interior. 
Thus, by making the initial torus marginally optically thick, we crudely mimic the model in which the torus is sustained against vertical collapse by the gradients of infrared radiation energy density \cite{KrolikInfrared}.
In what follows, we describe the torus outflow behavior at three different times after the initial  turn-on of X-rays.

After one orbital period, an over-pressured region extends  to 
$r\simeq 4\,{\rm pc}$
throughout the area that is not shadowed by the high density torus. At this time the non-shadowed area lies at $\theta<56^\circ$.
In this space, an axisymmetric region exists between $0.7{\rm pc}<\varpi<2.5{\rm  pc}$ and $z<7$ where the temperature is everywhere, of the order of the local virial temperature (except in the 
torus interior) :  $T\simeq 5\,T_{\rm vir}(r)$. 
Closer to the black hole, the heating is much stronger, leading to $T\sim 10^7 {\rm K}$.
High temperature, low density gas fills the space in the "throat" of the torus. 
The outer edge of the torus has extends to $\sim 5.3 \,{\rm pc}$ in temperature, and to $\sim 4.5 \,{\rm pc}$ in density contours. The ionization parameter in the same region
is $\xi\simeq 600$. A wind is well-developed: the velocity field in the vicinity of the high density torus, follows the equal pressure contours; the maximum radial velocity is observed close to the axes at $v_{\rm max}(\theta=3^\circ)=1180\,{\rm km}\,{\rm s}^{-1}$, the second local maximum is at $\theta=12^\circ$:  $v_{\rm max}=850\,{\rm km}\,{\rm s}^{-1}$. The maximum velocity
gradually decreases towards higher inclinations $v_{\rm max}(\theta=27^\circ)=343\,{\rm km}\,{\rm s}^{-1}$; and at $\theta=60^\circ$ it is $v_{\rm max}=110\,{\rm km}\,{\rm s}^{-1}$. 
Closer to black hole, at about $0.2\,{\rm pc}$, the $v_\theta$ component of the velocity is more pronounced, until heights $z\simeq \,1.4{\rm pc}$ are reached forming a base of the evaporative wind.  
After this time, the flow is approximately   symmetrical in both hemispheres. Because we are solving equations of ideal hydrodynamics (with only a small numerical viscosity), accretion through inner boundary (at r=0) is almost negligible:  $\dot{M_{\rm in}}(M_\odot\,{\rm yr}^{-1}) < 10^{-8}$. The mass-loss rate  
$\dot{M}_\Omega^{\rm max}(M_\odot\,{\rm yr}^{-1}\, {\rm sterrad}^{-1})$ peaks at  $\theta\simeq 10^\circ$, i.e. at much higher inclinations than $v_{\rm max}$. The total mass-loss rate at this stage is $\dot{M}(M_\odot\,{\rm yr}^{-1}) \simeq 3\cdot 10^{-4}$.

\begin{figure}
\includegraphics[width=640pt]{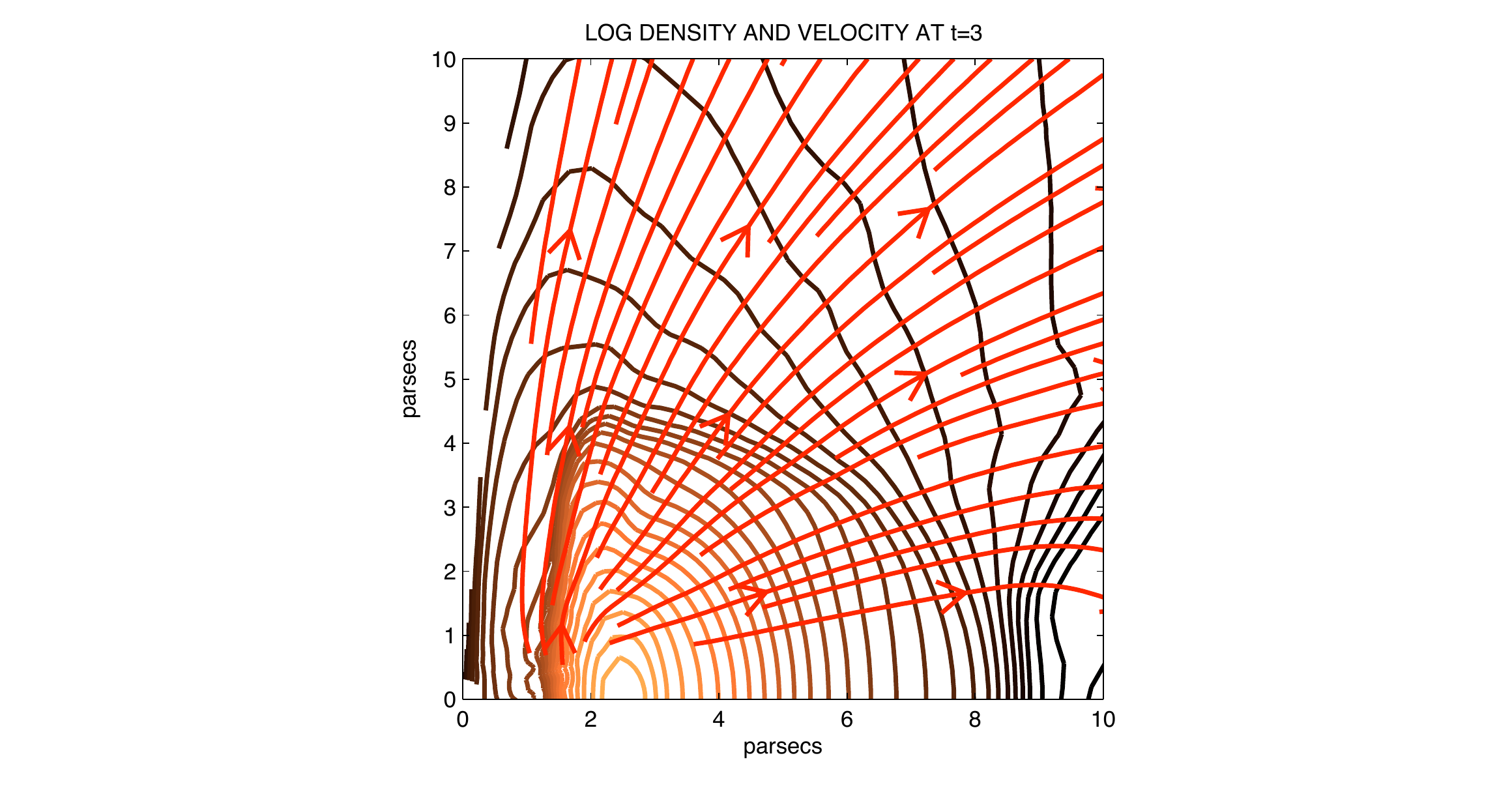}
\caption{Velocity streamlines superimposed on contours of  $\log(n/n_0)$
after three orbital periods. Note, that streamlines carry information about the direction only i.e. the magnitude of the velocity at low inclination is much higher than near the equatorial plane.} 
\label{Fig1_0}
\end{figure}

Figure \ref{Fig1_0} shows the flow streamlines and density contours at three orbital periods.  
At this time the high pressure region expands vertically, to $z\simeq 10 $ pc from the inner torus. The torus  
inner edge can be inferred from the temperature and density maps: $\varpi^-\simeq 1.3\,{\rm pc}$. Inside a throat of this torus,  
$T\simeq 10\,T_{\rm vir}$.  
The contours of $T$ become more flattened: $z_{\rm max}\simeq 4\,{\rm pc}$,  $\varpi^{+}\simeq 8\,{\rm pc}$.
The conservation of angular momentum helps to keep the outer edge of the torus at $\sim 5\,{\rm pc}$. However,
there is an envelope around this dense torus: $z_{\rm max}\simeq 4\,{\rm pc}$,  $\varpi^+\simeq 8.1\,{\rm pc}$.  
The values of the minimum ionization parameter and the column density vary significantly with the inclination 
angle: $\xi_{\rm min}(\theta=8^\circ)=2000$, $N_{22}=0.3$; at  $\theta \simeq 25^\circ$, the ionization parameter decreases significantly:
$\xi_{\rm min}(\theta=26^\circ)=63$, $N_{22}=2.3$; and at higher inclination, the ionization parameter is quite low: $\xi_{\rm min}(\theta=53^\circ)=6.6$, and the obscuration 
is quite high: $N_{22}=75$. The mass loss rate changes slowly $d\dot{M}/dt
((M_\odot\,{\rm yr}^{-1})/{\rm yr}) \simeq 1.53\cdot 10^{-6}$.
The mass-loss rate  peaks at  $\dot{M}^{\rm max}_\Omega=0.05$ at $\theta\simeq 20^\circ$, where  
$v_{\rm max}(\theta =20^\circ)=616\,{\rm km}\,{\rm s}^{-1}$; the maximum value of $v_{\rm max}(\theta =5^\circ)=1000\,{\rm km}\,{\rm s}^{-1}$; The total mass-loss rate is $\dot{M}(M_\odot\,{\rm yr}^{-1}) \simeq 0.08$. It is this evolved model that gives mass-loss rate in accord with the estimates of 
\cite{KrolikBegelman86}. 

\begin{figure}
\includegraphics[width=540pt]{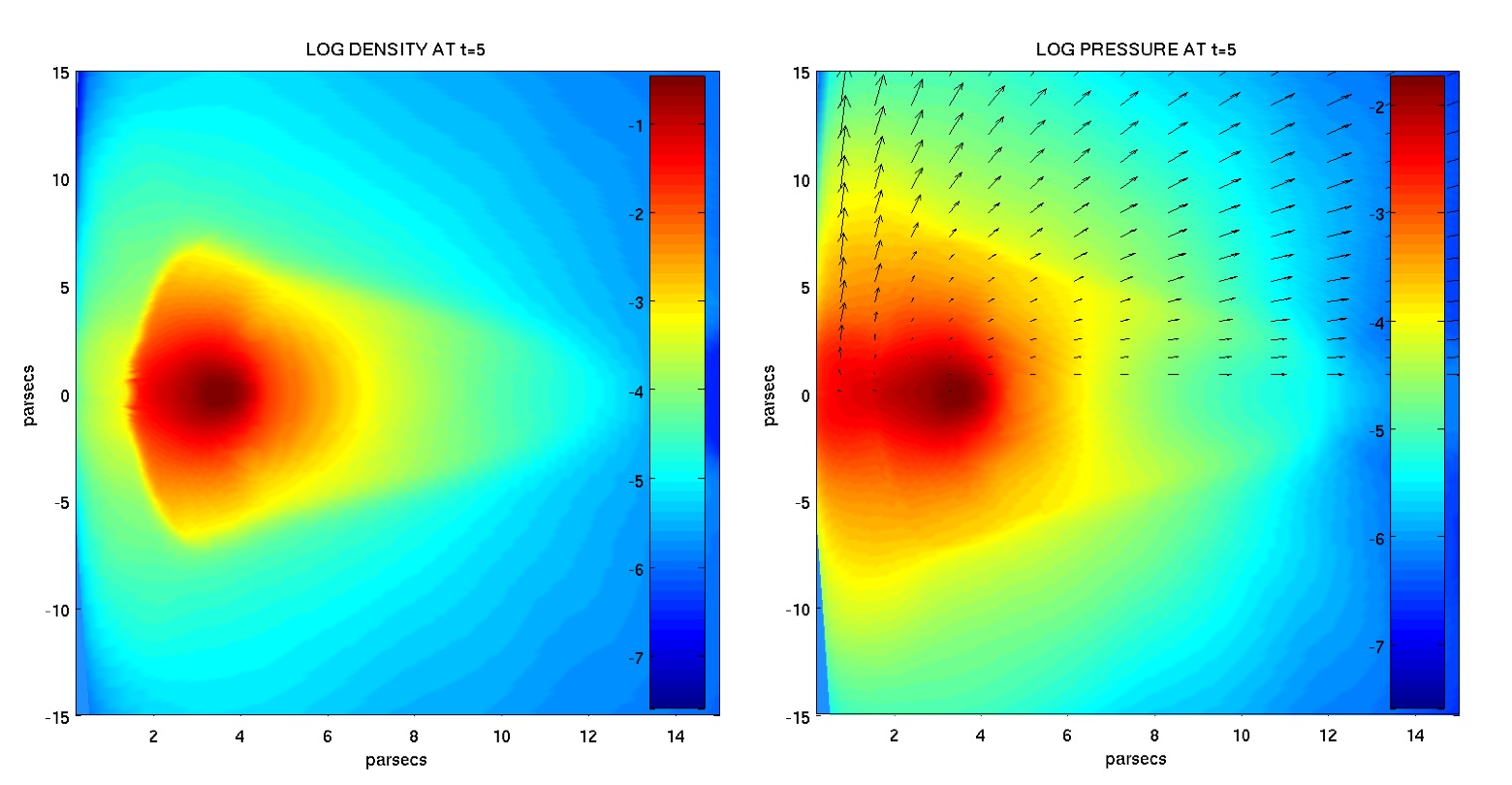}
\caption{Color-intensity plots of $\log(n/n_0)$ (left panel), and $\log(P/P_0)$ (right panel) at five orbital periods; at northern hemisphere superimposed with velocity vectors. }
\label{Fig2_0}
\end{figure}

Figure \ref{Fig2_0} shows the flow at five orbital periods.  At this time the density maximum is at $\varpi=3.6$.
Remarkably, the inner edge of the inner, dense torus is at the same distance from the black hole: $\varpi^-\simeq 1.3\,{\rm pc}$, as it was at three orbital periods, while the 
outer edge is at $\varpi^+\simeq 6.2\,{\rm pc}$.
The position of the outer edge of the torus, as infered from the temperature maps, extends to 11 pc. 
Outside this region the flow temperature is $T\simeq 2.5T_{\rm vir}$.
The significant drop of ionization parameter $\xi$ from $\sim 10^3$ to
$\sim 10$, occurs again at  $\theta =25^\circ$, where the 
column also rises to a $\sim \,10^{22}\rm$. Aspect ratio of the torus: $\Delta r/r\sim 1$ in accord with what is inferred from observations \cite{KrolikBegelman86, JaffeNATUR}.
Almost everywhere in the wind the poloidal component of the velocity is determmined by $v_r$.  However, 
closer to the black hole, the  $v_\theta$ component is important,  $v_\theta\sim v_r$.
The mass-loss rate is approximately constant at $15^\circ<\theta<60^\circ$: $\dot{M}^{\rm max}_\Omega=0.07$, and has a peak at $\theta=74^\circ$: $\dot{M}^{\rm max}_\Omega=0.16$. The total mass-loss rate is $\dot{M}(M_\odot\,{\rm yr}^{-1}) \simeq 0.25$ and $d\dot{M}/dt
((M_\odot\,{\rm yr}^{-1})/{\rm yr}) \simeq 10^{-6}$ .
The maximum value of $v_{\rm max}(\theta =3^\circ)=1032\,{\rm km}\,{\rm s}^{-1}$; another local peak is at $\theta=15^\circ$, where the maximum velocity is  
$v_{\rm max}=606\,{\rm km}\,{\rm s}^{-1}$, and the minimum $v_{\rm max}(\theta =90^\circ)=180\,{\rm km}\,{\rm s}^{-1}$. The torus is losing mass in all directions, although with very different speed at different inclinations.

\begin{figure}
\includegraphics[width=440pt]{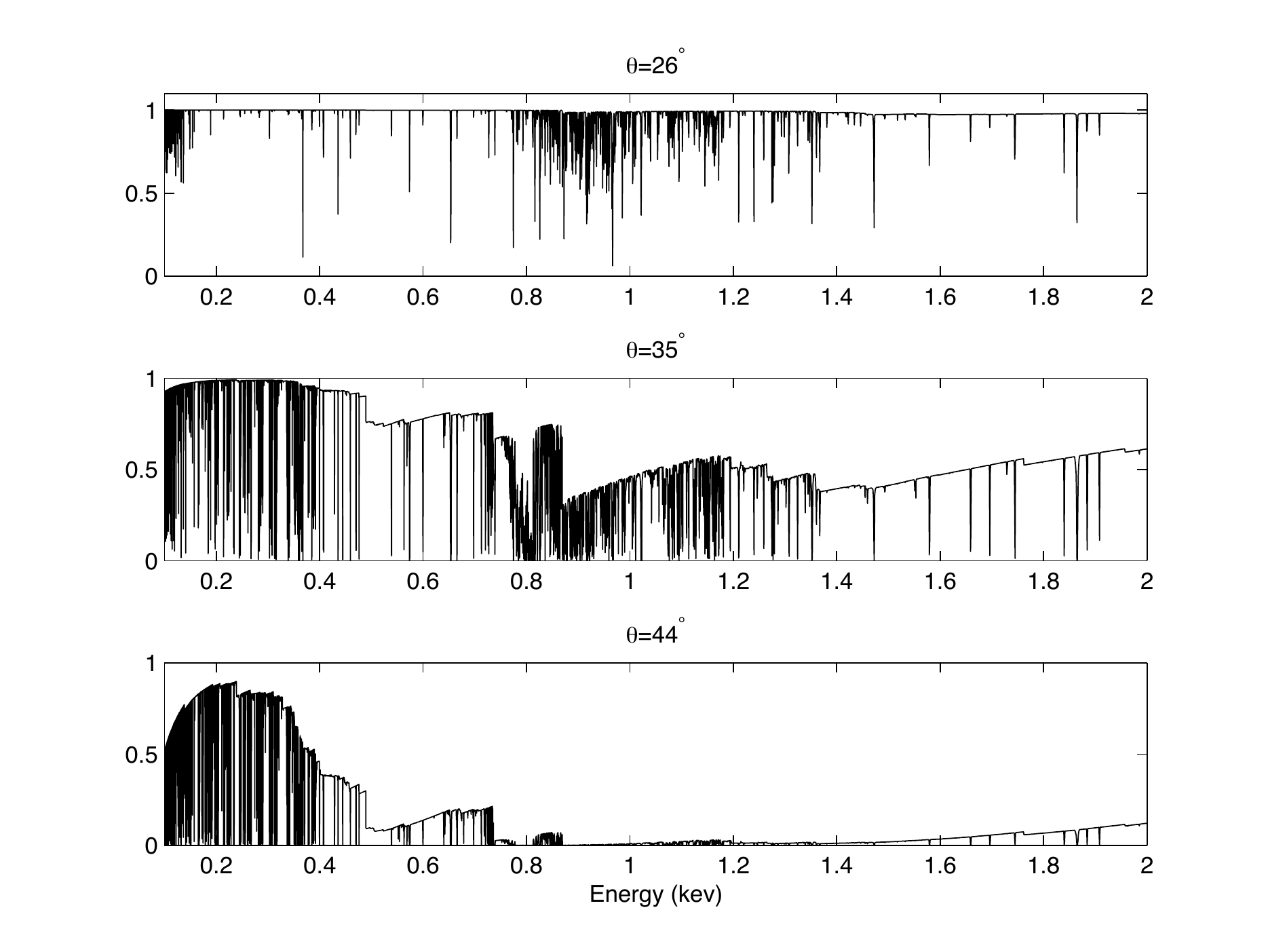}
\caption{The transmitted spectra observed at three orbital periods from 2.5D model of the evaporative torus wind. Each panel corresponds to a 
different viewing angle, $\theta=26^\circ,\,35^\circ,\,44^\circ$; x-axis: energy; y-axis: normalized, transmitted flux. The initial spectrum is a power law with energy index $\alpha=1$.} 
\label{Fig3_0}

\end{figure}

We  have performed multidimensional simulations of a wind  excited by X-ray heating of a rotationally supported torus. 
Our dynamical calculations predict the velocity field and the spatial distribution of opacity and emissivity of the gas in 
both continuum and lines as a function of position. 
( Synthesizing the spectrum provides a key test for our model ).  We have calculated synthetic spectra 
by integrating over the wind and using the formal solution of the transfer equation.
This is done by passing the results of the hydrodynamical calculations into a code which calls the subroutines 
from the {\sc xstar} code \cite{KallmanBautista01} to calculate the transmitted flux. 
In this paper we present only absorption spectra. This approximately represents the case when warm absorber gas lies on 
the line of sight to the observer and we see only transmitted spectra.  These are shown 
in Figure \ref{Fig3_0}, {for various angles $\theta=26^\circ,\,35^\circ,\,44^\circ$ at three orbital periods}.   These show that 
our simulations provide gas with intermediate ionization and sufficient column to produce 
observable warm absorber features.  
This supports the idea that it is this wind that is responsible for the warm absorber gas observed in  Seyfert galaxies.
More detailed results, including dependence of these results on different parameters as well as limitations of the adopted model  
and comparisons with observations, will be discussed in a separate paper.

This research was supported by an appointment to the NASA Postdoctoral Program at the NASA Goddard Space Flight Center, administered by Oak Ridge Associated Universities through a contract with NASA, and by grants from the NASA Astrophysics Theory Program 05-ATP05-18.

\clearpage

\end{document}